\documentclass[
    aps,
    prb,
    reprint,
    superscriptaddress,
    amsmath,
    amssymb
]{revtex4-2}
\usepackage{graphicx}
\usepackage{placeins}
\usepackage{color}
\usepackage{dcolumn}
\usepackage{latexsym}

\usepackage[normalem]{ulem}
\usepackage{hyperref,amssymb}
\usepackage{url}
\usepackage{color,soul}
\usepackage{graphicx}
\usepackage{verbatim}
\usepackage{multirow}
\usepackage{amsmath}
\newcommand{\beq}{\begin{eqnarray}}
\newcommand{\eeq}{\end{eqnarray}}
\usepackage{mathrsfs}
\usepackage{float,soul}
\usepackage[dvipsnames]{xcolor}
\usepackage{mathtools}
\usepackage{slashed}
\usepackage{physics}
\usepackage{graphicx}
\usepackage{epstopdf}
\usepackage{subfigure}
\usepackage[font=small]{caption}
\usepackage{hyperref}
\usepackage{bbold}
\usepackage{wasysym}
\usepackage{feynmp}
\usepackage{hyperref}
\hypersetup{colorlinks}
\usepackage{ragged2e}
\usepackage{siunitx}
\usepackage[utf8]{inputenc}
\usepackage{bm}

\renewcommand{\vec}[1]{\bm{#1}}

\begin{document}

\title{2D Transport in an in-plane magnetic field}
\author{Aryaman Babbar}
\author{Sankar Das Sarma}
\affiliation{Condensed Matter Theory Center and Joint Quantum Institute, Department of Physics, University of Maryland, College Park, Maryland 20742, USA}

\begin{abstract}
A parallel in-plane magnetic field could, in principle, distinguish between two competing physical scenarios for the experimentally observed density-tuned 2D metal-insulator transition (where decreasing the carrier density leads to a crossover from an effective metal to an effective insulator): Wigner crystallization or Anderson localization. Importantly, the observation of such a density-driven metal-insulator transition alone does not uniquely determine the microscopic nature of the insulating state. Since the main scattering mechanism in 2D doped semiconductors arises from screened random charged impurities and screening in turn depends on the electronic density of states, the in-plane magnetic field could distinguish between the two by decreasing screening through spin polarization and this enhances the effective critical density for Anderson localization compared with Wigner crystallization. We give the general theory and  provide results for the quantitative magnitudes of the spin polarization effect on the transition density by focusing on two recent experiments [Z. Ge, et al, Nature (London) 654, 902 (2026), T. Han, et al, arXiv:2604.00113], noting that the critical density may actually decrease if the dominant scattering is by short-ranged defects instead of long-ranged charged impurities. The difference between the two cases arises from whether spin polarization dominates screening (enhanced critical density) or the Fermi surface (suppressed critical density).
\end{abstract}

\maketitle

\section{Introduction}

In 1934, Eugene Wigner pointed out that an interacting electron gas in a neutralizing positively charged jellium background would crystallize into a ``Wigner crystal" (WC) at low electron densities at $T=0$ (in contrast to classical systems which are always solids at $T=0$) to minimize its Coulomb energy, which dominates over the quantum zero-point kinetic energy~\cite{wigner1934on}. Realizing a WC in the laboratory has been a persistent theme in condensed matter physics for $> 50$ years~\cite{fertig1996properties,shayegan1996case,monarkaha20122D,shayegan2022wigner,ke2025exploring}.

2D semiconductors have been the most studied system in the context of realizing Wigner crystals, perhaps because all 3D metals are electron liquids with their zero point quantum  kinetic energy (i.e. the Fermi energy) being very high $\left(T_F> 10^4 \text{ K}\right)$ by virtue of their high electron density, thus making a WC a rather exotic object. These doped (two-dimensional) materials allow a continuous change of the charge carrier density through gating, so the density-dependent properties of the system can be studied in a single sample to see whether a WC forms at very low densities~\cite{fertig1996properties,shayegan1996case,monarkaha20122D,shayegan2022wigner,ke2025exploring,ando1982review}. A quantum WC forms when the typical potential energy $V$ dominates over the typical kinetic energy $K$, i.e., $V \gg K$. The Wigner-Seitz radius $r_S$ is a dimensionless measure of the ratio of the typical $V$ and $K$ values, given by $r_S = m e^2/\left(\kappa \hbar^2 \sqrt{\pi n}\right)$ in 2D, where $m$ is the effective mass of the charge carriers, $\kappa$ is the relative permittivity and $n$ is the charge carrier density. The threshold above which a 2D WC is formed is quantified by Quantum Monte Carlo (QMC) simulations. QMC simulations have shown that in a clean, disorder-free system, the WC transition is observed at $r_S \approx 37$~\cite{drummond2009phase, 2DEG_GS_QMC_WC2, diffusion_MC_QMC_WC3, vinko_QMC_WC4, flatiron_QMC_WC5}. The electron density corresponding to $r_S \approx 37$ is $n_\text{WC}$. For $n< (>) \, n_\text{WC}$, the 2D system is a WC (electron liquid) at $T=0$.

Most experiments attempting to realize a WC in the lab are transport experiments on these 2D semiconductors, where the sample, with a fixed disorder configuration, makes a metal-insulator transition (MIT) from an effective ``good metal" (resistivity increasing with temperature at low temperatures) to an effective ``good insulator" (resistivity decreasing with temperature at low temperatures) at a certain critical charge carrier density $n_c$~\cite{ahn2023density,dassarma2014two}. If this measured $n_c$ turns out to be close to $n_\text{WC}$, then a WC discovery is claimed and the MIT is then claimed to be arising from Wigner crystallization. We note that both metallic (insulating) phases for $n > (<) \, n_c$ are `effective' low temperature phases since the measurements are necessarily not at $T=0$, and the MIT is actually a metal-to-insulator crossover for $n \sim n_c$ rather than a sharp transition.

However, associating such an effective MIT with Wigner crystallization comes with a few issues. WC is not an insulator unless it is strongly pinned due to impurities in the sample -- in that sense, the MIT is really due to the formation of a pinned WC -- with no disorder, the WC is a metal. Additionally, the 2D semiconductor samples inevitably have random quenched charged (long-range) and isovalent (short-range) impurities. Therefore, even for the so-called ultra-high quality systems~\cite{ultra_high_mobility_theory, ultra_high_mobility_experiment}, if $n$ is reduced to form a WC in a sample, the ratio of the impurities to charge carrier density $n_i/n$ also increases, and in such a case, the effect of disorder on the system must also be considered, since the process of reducing carrier density to enhance $r_S$ (and hence, interaction strength) necessarily also enhances the effective disorder strength by enhancing $n_i/n$~\cite{ahn2023density}. There have been experiments in extremely low disorder 2D systems \cite{manfraetal}, where $n$ has been reduced to well below $n_\text{WC}$ (in fact increasing $r_S$ to $>50$!) and even then an MIT is not observed -- the MIT is observed at a much lower density than $n_\text{WC}$ in this extreme high-quality sample, so this transport experiment is consistent with a disorder-driven transition. In fact, most MITs happen at an experimental density $n_\text{MIT}$ with $n_\text{MIT} \sim n_i$ ~\cite{ahn2023density}, emphasizing the importance of disorder in an MIT that is claimed to arise from the formation of a pinned WC. One should, therefore, use extreme caution in identifying the formation of a WC just by virtue of the 2D MIT crossover density $n_c$ being comparable with the putative theoretical WC formation density $n_\text{WC}$, as is universally done by the experimentalists.

The transport experiments cannot distinguish between the mechanisms of MITs. In fact, transport showing an effective 2D MIT is simply a localization transition for the electrons as they crossover from an effective high-density metallic extended state to an effective low-density localized insulating state with no information on the underlying insulating phase being a WC just based on transport measurements. We need to analyze the system and check whether the density $n_c$ at which the transition happens is more consistent with a pinned WC transition, or with a disorder-driven strong localization transition. For the case where we have only charged impurities in the system, these charged impurities are screened by the charge carriers in the system, therefore, each charge carrier experiences an effective potential due to these charged impurities. This can be used to formulate a single particle potential experienced by a single charged particle by using the Random Phase Approximation. We can calculate the mean free path $l$ of a single charge carrier under this single particle potential using Boltzmann theory, and the $n_c = n_\text{IRM}$ at which the MIT happens is given by the Ioffe-Regel-Mott (IRM) criterion for strong localization, $k_F \left(n_c\right) l \left(n_c\right) \sim 1$~\cite{IoffeRegel1960, Mott1974, MottDavis1979}. This then provides an alternative scenario where this IRM $n_c$ should be compared with the experimental $n_\text{MIT}$ in addition to just comparing $n_\text{MIT}$ with the theoretical $n_\text{WC}$~\cite{ahn2023density}.

\begin{figure}
    \centering
    \includegraphics[width=0.9\linewidth]{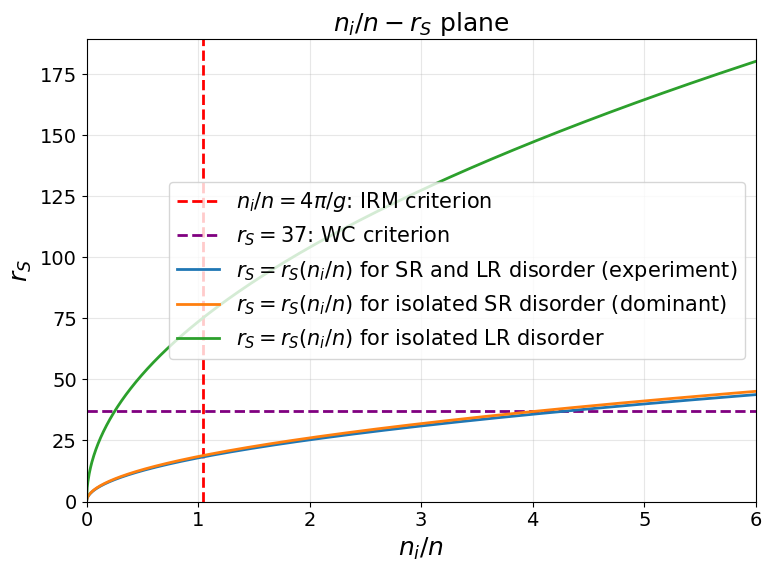}
    \caption{The horizontal and vertical lines represent the WC and IRM criterion respectively. We plot $r_S$ as a function of $n_i/n$ for the experiment~\cite{ge2025visualizing, haleemprivatecomm}. The various curves in the figure indicate how changing carrier density in a sample tunes interaction $(r_S)$ and disorder $(n_i/n)$ with the top green curve indicating the highest quality sample and the lower curves (blue and orange) indicating lower quality samples of higher impurity density.}
    \label{fig:WC_AL}
\end{figure}

We can schematically represent the effective `phase diagram' in the $n_i/n - r_S$ plane, where $n_i$ provides an effective measure of the combined density and strength of disorder. For each disorder source $D$, we define $n_D$ through its contribution to the scattering rate at $T = 0$, $1/\tau_D=\left(\hbar/m\right)n_D$, and write $n_i=\sum_D n_D$. In the system in~\cite{ge2025visualizing, haleemprivatecomm}, $n_i$ is approximately independent of the carrier density $n$, since the disorder arises primarily from short range defects and not from long range random charged impurities. Even the charged impurities in the sample are strongly screened, making their contribution to $n_i$ approximately independent of the carrier density. The approximate criterion for the formation of a WC is $r_S>37$, whereas the IRM criterion becomes
$n_i/n > 4\pi/g$, where $g$ is the combined spin and valley degeneracy~\cite{babbar2026wignersolidandersonsolid}. These two boundaries are shown in Figure~\ref{fig:WC_AL}. As the carrier density is reduced, the system reaches the WC criterion before the IRM criterion only if the disorder is sufficiently weak. Quantitatively, this requires $n_i< \frac{4\pi}{g}
\left(\frac{m e^2} {\kappa\hbar^2 \left(37\sqrt{\pi}\right)}
\right)^2$. To the best of our knowledge, this extreme high purity was achieved only once in an MIT experiment in~\cite{manfraetal}, where the MIT was interpreted as a localization transition and not as a WC transition because $n_\text{MIT} \ll n_\text{WC}$.

It is also useful to express $r_S$ directly as a function of $n_i/n$. Since $r_S\propto n^{-1/2}$ while $n_i/n\propto n^{-1}$, we have that $n_i/n\propto r_S^2$. Thus, as the carrier density is reduced, $n_i/n$ grows quadratically with $r_S$. We plot this trajectory for the experimentally relevant case containing both short- and long-range disorder, as well as for the limiting cases in which the two disorder contributions are considered separately. The resulting curves show that a density-tuned system moves toward the IRM boundary more rapidly than toward the WC boundary. This further emphasizes that disorder cannot be neglected when interpreting an MIT as the formation of a pinned WC. The key point, which is often missed in the 2D WC discussion, is that increasing (decreasing) $r_S$ $(n)$ by itself without affecting disorder is impossible in a given sample since increasing $r_S$ automatically increases the dimensionless disorder $n_i/n$, which is incorporated in the phase diagram of Figure \ref{fig:WC_AL}.

Due to screening of the charged impurities by the carriers themselves, the value of $n_c$ is dependent on the total degeneracy (spin and valley) of the charge carriers, $g$, because screening is proportional to the density of states, and hence to $g$. On the other hand, $r_S$ is independent of the degeneracy of the system, as it is simply a dimensionless measure of the inter-particle separation. If we apply a parallel magnetic field to the system, the only effect it has is to change the energies of different spins of the charge carriers through the Zeeman coupling since a parallel in-plane field does not couple to the orbital motion. In this work, we assume that the in-plane field only couples to the spin through the Zeeman coupling, thus, modifying the spin degeneracy, but without affecting the 2D orbital motion and the valley degeneracy. If this field is strong enough, the charge carriers can be polarized into a single spin configuration, changing the degeneracy $g$. This means that if the transition is due to disorder-driven localization, the $n_c$ will change when the field is applied, but if the transition is due to the formation of a pinned WC, $n_c$ does not change. In that sense, a parallel magnetic field can be used to distinguish between a pinned WC transition and a strong localization driven transition.

Recently, there have been experiments on bilayer MoSe$_2$ at Berkeley~\cite{ge2025visualizing, haleemprivatecomm} and rhombohedral graphene at MIT~\cite{longjumetallicwignercrystal} that claim the realization of Wigner crystals in 2D semiconductor systems by lowering the charge carrier density. In the section below, inspired by these two experiments, we calculate how the density $n_c$ at which the Metal-Insulator transition happens changes, when we apply a parallel magnetic field, if these transitions are due to disorder-driven localization. We calculate the temperature dependent magneto-resistance in the metallic regime using RPA-Boltzmann theory that is valid in the metallic regime. We also argue that the degeneracy $g$ that inevitably enters the IRM criterion can be given a suitable value $g_\text{eff}$ at intermediate magnetic field values, and using this value of $g_\text{eff}$, we can calculate the $n_c$ at which the transition happens at different field values.

\section{Calculation}

We consider a 2D electron system, with electron density $n$, parabolic dispersion characterized by an effective mass $m$, and valley degeneracy $g_v$ (which we assume to be unaffected by the applied magnetic field). The electrons have a spin degeneracy of $2$, which is lifted when a parallel magnetic field is applied, since the applied field couples to the spin through Zeeman effect -- there is no magneto-orbital effect because the applied field is parallel to the 2D layer. We assume that the only effect of this magnetic field is to increase the number of electrons that are spin-down, and reduce the number of electrons that are spin-up. The Land\'e $g$ factor is denoted by $g_L$.

The energy of electrons with $\sigma = \pm$, where $\sigma$ denotes the direction of the spin with reference to $B$, with wave vector $\vec{k}$ is given by

\begin{equation}
    E = \frac{\hbar^2 k^2}{2m} + \sigma \frac{g_L \mu_B B}{2}.
\end{equation}

Here, $\sigma = +$ means that the spin is in the direction of the magnetic field (spin-up) and $\sigma = -$ means that the spin is opposite to the magnetic field (spin-down). In a magnetic field, a spin-up $\left(\sigma = +\right)$ electron has its magnetic moment antiparallel to the field, so its Zeeman energy is increased. Conversely, for a spin-down $\left(\sigma = -\right)$ electron, the magnetic moment is parallel to the field and the Zeeman energy is lowered. We can absorb this $\pm g_L \mu_B B/2$ into the chemical potential by saying that the chemical potential of the $+$ spin, $\mu_+$ is distinct from that of the $-$ spin, $\mu_-$. The two chemical potentials differ by a known $g_L \mu_B B$, as $\mu_- - \mu_+ = g_L \mu_B B$. The number density 

\begin{equation}
    n_\pm = g_v \int \frac{\text{d}^2 k}{\left(2 \pi\right)^2} \frac{1}{\exp{\left(\frac{1}{k_B T} \left(\frac{\hbar^2 k^2}{2m} - \mu_\pm\right)\right)}+1}. 
\end{equation}

This leads to

\begin{equation} \label{eq:n_pm}
    n_\pm = \frac{m g_v k_B T}{2 \pi \hbar^2} \ln{\left(1 + e^{\beta \mu_\pm}\right)}. 
\end{equation}

$n = n_+ + n_-$, which is the experimental parameter we can control, so substituting $\mu_\pm = k_B T \ln{\left(x\right)} \mp g_L \mu_B B/2$ yields

\begin{equation}
    n = \frac{m g_v k_B T}{2 \pi \hbar^2} \ln{\left(1 + x^2 + 2x \cosh{\left(\frac{g_L \mu_B B}{2 k_B T}\right)}\right)}.
\end{equation}

Solving for $x$, and taking the root that reproduces the $B = 0$ result gives

\begin{equation}
\begin{aligned}
    x =& \sqrt{\sinh^2{\left(\frac{g_L \mu_B B}{2 k_B T}\right)} + \exp{\left(\frac{2 \pi \hbar^2 n}{g_v m k_B T}\right)}} \\
    &- \cosh{\left(\frac{g_L \mu_B B}{2 k_B T}\right)}.
\end{aligned}
\end{equation}

Therefore, 

\begin{equation}
\label{eq:mu_pm}
\begin{aligned}
\mu_\pm
={}& k_B T
\ln\left[
    \sqrt{
        \sinh^2\!\left(
            \frac{g_L\mu_B B}{2k_B T}
        \right)
        +
        \exp\!\left(
            \frac{2\pi\hbar^2 n}{g_v m k_B T}
        \right)
    }
    \right. \\
&\left.
    -\cosh\!\left(
        \frac{g_L\mu_B B}{2k_B T}
    \right)
\right]
\mp \frac{g_L\mu_B B}{2}.
\end{aligned}
\end{equation}

With this, we can think about the polarizability in a system of interacting electrons. We suppose that the magnetic field parallel to the plane of the system only interacts with the spin of the electrons as presented above. The polarizability at a wavevector $\vec{q}$, frequency $\omega$ and temperature $T$ is given by, for a 2D system of area $A$~\cite{bruus_flensberg_2004}

\begin{equation}
    \frac{1}{A} \sum_{g_v} \sum_{\sigma = \pm} \sum_{\vec{k}} \frac{n_{\vec{k} + \vec{q}}^\sigma - n_{\vec{k}}^\sigma}{\omega - \epsilon_{\vec{k} + \vec{q}} + \epsilon_{\vec{k}}},
\end{equation}

where $\epsilon_{\vec{k}} = \hbar^2 k^2/\left(2m\right)$ and $n_{\vec{k}}^\sigma$ is the Fermi occupation number of spin $\sigma$ and electron wavevector $\vec{k}$.

Converting the sum over $\vec{k}$ to an integral, we get

\begin{equation}
    \Pi \left(\vec{q}, \omega, T, \{\mu_\sigma\}\right) = g_v \sum_\sigma \int \frac{\text{d}^2 k}{\left(2 \pi\right)^2} \frac{n_{\vec{k} + \vec{q}}^{\sigma} - n_{\vec{k}}^{\sigma}}{\omega - \epsilon_{\vec{k} + \vec{q}} + \epsilon_{\vec{k}}}.
\end{equation}

We plug 

\begin{equation}
    n_{\vec{k}} =
    \int_0^{\infty} \text{d} \mu' \frac{\Theta\left(\mu' - \epsilon_{\vec{k}}\right)}
    {4 T \cosh^2\left(\frac{\mu_\sigma - \mu'}{2 k_B T}\right)},
\end{equation}

to get

\begin{equation}
\begin{aligned}
\Pi\left(\vec{q},\omega=0,T,\{\mu_\sigma\}\right)
={}&
\sum_\sigma \int_0^\infty \mathrm{d}\mu'\,
\\
&
\frac{
    \Pi_\sigma\left(\vec{q},\omega=0,T=0,\mu'\right)
}{
    4T\cosh^2\!\left(
        \dfrac{\mu_\sigma-\mu'}{2k_B T}
    \right)
}.
\end{aligned}
\end{equation}

where after simplifying

\begin{equation}
\begin{aligned}
\Pi_\sigma\left(\vec{q},\omega=0,T=0,\mu_\sigma\right)
={}&
\frac{g_v m}{2\pi\hbar^2}
\Biggl[
1-
\sqrt{
    1-\frac{8m\mu_\sigma}{\hbar^2q^2}
}
\\
&\qquad\qquad\times
\Theta\left(
    q^2-\frac{8m\mu_\sigma}{\hbar^2}
\right)
\Biggr].
\end{aligned}
\end{equation}

We use this polarizability when we discuss the calculation of scattering rates under the RPA-Boltzmann approximation. Note that since we are discussing DC transport, only the static zero frequency polarizability is relevant for our consideration.

If we consider scattering of electrons by a charged impurity of charge $e$, then the transport scattering rate for a layer of charged impurities at perpendicular distance $z$ from the plane of the material with number density $n_i$, $\tau_t^{-1}$ is given by~\cite{yihuangsds} 

\begin{equation} \label{eq:charged_scattering_rate}
\begin{aligned}
    &\frac{1}{\tau_t \left(k\right)}
    =
    \frac{4 m n_i}{\pi \hbar^3}
    \int_0^{\pi/2} \mathrm{d}\theta \,
    \sin^2{\left(\theta\right)} \\
    &\left(\frac{
        \frac{2 \pi e^2}{\kappa \left(2 k \sin{\left(\theta\right)}\right)}
        e^{-2 k \sin{\left(\theta\right)} z}
    }{
        1
        +
        \Pi \left(2k \sin{\left(\theta\right)}, \omega = 0, T, \{\mu_\sigma\}\right)
        \frac{2 \pi e^2}{\kappa \left(2 k \sin{\left(\theta\right)}\right)}
    }\right)^2.
\end{aligned}    
\end{equation}

Note that this scattering rate is the same for both spins $\sigma = \pm$. Therefore, $\tau_{+, t} \left(k\right) = \tau_{-, t} \left(k\right) = \tau_{t} \left(k\right)$. We can also compute the scattering rate due to short-ranged impurities in the system similarly as in~\cite{exponentspaper} because it does not depend on the polarizability. In writing Eq. (\ref{eq:charged_scattering_rate}), we use the fact that the charged impurity is screened by the electrons themselves resulting in the denominator on the right hand side of Eq. (\ref{eq:charged_scattering_rate}) involving the polarizability.  Note that we are specifically considering monovalent charged impurities, but taking into account multivalent impurities will only introduce a factor of impurity charge, which we absorb into $n_i$.

At a particular $\vec{k}$, the scattering rates add, so $\tau_{t, \text{tot}}^{-1} = \sum_i \tau_{t,i}^{-1}$, where $\tau_{t,i}^{-1}$ are the scattering rates due to different mechanisms.

We consider the Boltzmann kinetic equation in the presence of a weak electric field $\vec{E}$ for electrons along spin $\sigma$~\cite{yihuangsds}. The equilibrium distribution function (for when $\vec{E} = \vec{0}$) is given by $n_{0, \sigma} \left(\vec{r}, \vec{p}\right)$ and the perturbed distribution function is $n_\sigma \left(\vec{r}, \vec{p}\right) = n_{0, \sigma} \left(\vec{r}, \vec{p}\right) + n_{1, \sigma} \left(\vec{r}, \vec{p}\right)$. The distribution function is normalized such that

\begin{equation}
    g_v \int \frac{\text{d}^2 p}{\left(2 \pi \hbar\right)^2} \int \text{d}^2 r \, \,  n_\sigma \left(\vec{r}, \vec{p}\right) = N_\sigma,
\end{equation}

where $N_\sigma$ is the number of electrons in the system with spin $\sigma$. The Boltzmann kinetic equation is

\begin{equation}
    \frac{\partial n_\sigma}{\partial t} + \vec{v} \cdot \frac{\partial n_\sigma}{\partial \vec{r}} - e \vec{E} \cdot \frac{\partial n_\sigma}{\partial \vec{p}} = - \frac{n_\sigma - n_{0, \sigma}}{\tau_{t, \sigma} \left(k\right)}.
\end{equation}

Assuming $E$ and $n_1$ are of the same order in some small parameter, we consider the equation up to that order by assuming that $n_1$ does not depend on $t$ and $\vec{r}$. This gives 

\begin{equation}
    - e \vec{E} \cdot \frac{\partial n_{0 ,\sigma}}{\partial \vec{p}} = - \frac{n_{1, \sigma}}{\tau_{t, \sigma} \left(\vec{k}\right)}.
\end{equation}

The current density $\vec{J}_\sigma$ due to electrons with spin $\sigma$ is given by

\begin{equation}
    \vec{J}_\sigma = - g_v e \int \frac{\text{d}^{2} p}{\left(2 \pi \hbar\right)^2} \vec{v} \,  n_{1, \sigma} \left(\vec{r}, \vec{p}\right). 
\end{equation}

We obtain with our normalization conventions

\begin{equation}
    n_{0, \sigma} \left(\vec{r}, \vec{p}\right) = n_{\vec{k}}^\sigma, \text{ where } \vec{p} = \hbar \vec{k} .
\end{equation}

Therefore, by doing a change of variables and by noting that $\frac{\partial n_{0 ,\sigma}}{\partial \vec{p}} = \vec{v} \frac{\partial n_{0 ,\sigma}}{\partial \epsilon}$, where $\epsilon = p^2/\left(2 m\right)$, we get

\begin{equation}
    \vec{J}_\sigma =  - e^2 \iint \text{d} \epsilon \, \text{d} \Omega \,  \vec{v}  \left(\vec{v} \cdot \vec{E}\right) \frac{\partial n^\sigma \left(\epsilon\right)}{\partial\epsilon} \, \nu_\sigma \left(\epsilon\right) \frac{1}{\Omega_0} \tau_{t, \sigma} \left(\epsilon\right).
\end{equation}

In this equation, the angular integration is over the direction $\hat{\Omega}$, and this direction is the same as that of $\vec{v}$ and $v = \sqrt{2E/m}$. $n^\sigma \left(\epsilon\right) = \left(\exp{\left(\beta \left(\epsilon - \mu_\sigma\right)\right)} + 1\right)^{-1}$ and $\Omega_0 = 2 \pi$ for our case. $\nu_\sigma \left(\epsilon\right)$ is the density of states with spin $\sigma$ and $\nu_\sigma \left(\epsilon\right) = \frac{m g_v}{2\pi\hbar^2}$. $\tau_{t, \sigma} \left(\epsilon\right)$ is the scattering time at $k$ corresponding to energy $\epsilon$. 

This gives

\begin{equation}
    \vec{J}_\sigma =  \frac{e^2}{m} \int \text{d} \epsilon \,  \epsilon\left(-\frac{\partial n^\sigma \left(\epsilon\right)}{\partial\epsilon}\right) \, \nu_\sigma \left(\epsilon\right) \tau_{t} \left(\epsilon\right) \vec{E}.
\end{equation}

Since $n_\sigma = \int \text{d} \epsilon \,  \epsilon\left(-\frac{\partial n^\sigma \left(\epsilon\right)}{\partial\epsilon}\right) \, \nu_\sigma \left(\epsilon\right)$, and the current densities due to the two spins add,

\begin{equation}
\begin{aligned}
\vec{J}
={}&
\frac{n e^2}{m}
\frac{
\begin{aligned}
&\Biggl[\int \mathrm{d}\epsilon\,
\epsilon
\left(
-\frac{\partial n^+(\epsilon)}{\partial\epsilon}
\right)
\nu_+(\epsilon)\tau_t(\epsilon)
\\
&\quad+
\int \mathrm{d}\epsilon\,
\epsilon
\left(
-\frac{\partial n^-(\epsilon)}{\partial\epsilon}
\right)
\nu_-(\epsilon)\tau_t(\epsilon) \Biggr]
\end{aligned}
}{
\begin{aligned}
&\Biggl[\int \mathrm{d}\epsilon\,
\epsilon
\left(
-\frac{\partial n^+(\epsilon)}{\partial\epsilon}
\right)
\nu_+(\epsilon)
\\
&\quad+
\int \mathrm{d}\epsilon\,
\epsilon
\left(
-\frac{\partial n^-(\epsilon)}{\partial\epsilon}
\right)
\nu_-(\epsilon)
\Biggr]\end{aligned}
}
\vec{E}.
\end{aligned}
\end{equation}

This means that the thermal and spin averaged scattering time is given by

\begin{equation}
    \tau_T = \frac{\begin{aligned}
&\Biggl[\int \mathrm{d}\epsilon\,
\epsilon
\left(
-\frac{\partial n^+(\epsilon)}{\partial\epsilon}
\right)
\nu_+(\epsilon)\tau_t(\epsilon)
\\
&\quad+
\int \mathrm{d}\epsilon\,
\epsilon
\left(
-\frac{\partial n^-(\epsilon)}{\partial\epsilon}
\right)
\nu_-(\epsilon)\tau_t(\epsilon) \Biggr]
\end{aligned}
}{
\begin{aligned}
&\Biggl[\int \mathrm{d}\epsilon\,
\epsilon
\left(
-\frac{\partial n^+(\epsilon)}{\partial\epsilon}
\right)
\nu_+(\epsilon)
\\
&\quad+
\int \mathrm{d}\epsilon\,
\epsilon
\left(
-\frac{\partial n^-(\epsilon)}{\partial\epsilon}
\right)
\nu_-(\epsilon)
\Biggr]\end{aligned}}
\end{equation}

The resistivity is given by $\rho = m/\left(n e^2 \tau_T\right)$.

The Ioffe-Regel-Mott (IRM) criterion for strong localization is given by $\rho = \left(2/g\right) \left(h/e^2\right)$, but here, while we have computed $\rho$, it is unclear what $g$, the total degeneracy is. We note that the IRM criterion is for $T = 0$. At $T = 0$, we have 

\begin{equation}
    \tau_0 = \frac{n_+ \tau_+ + n_- \tau_-}{n},
\end{equation}

where $n_\pm$ are the number densities of electrons with $\pm$ spins. The $B$ value required to polarize the entire sample is $B_c = \left(2 \pi \hbar^2 n\right)/\left(g_v m g_L \mu_B\right)$. The spin-polarized densities are: $n_{\pm} = n/2 \left(1 \mp B/B_c\right)$ for $B < B_c$ and $n_- = n, n_+ = 0$ for $B > B_c$. For $B > B_c$, $g_\text{eff} = g_v$. Note that $B_c = 2 E_F/\left(g_L \mu_B\right)$, where putting $g_L = 2$, we have $B_c = E_F/\mu_B$, where $E_F$ is the Fermi energy at $B = 0$.

Below we derive an equation for $g_\text{eff} \left(B\right)$ for arbitrary $B$ upto $B_c$. We have $k_F^{\pm} = \sqrt{4 \pi n_{\pm}/g_v} = \sqrt{2 \pi n/g_v \left(1 \mp B/B_c\right)}$. The polarizability at $T = 0$ is given by

\begin{equation}
\begin{aligned}
    \Pi_0 \left(q\right) = \frac{g_v m}{2 \pi \hbar^2} \Biggl(&2 - \sqrt{1 - \left(\frac{2 k_F^+}{q}\right)^2} \Theta \left(q - 2 k_F^+\right) \\
    &- \sqrt{1 - \left(\frac{2 k_F^-}{q}\right)^2} \Theta \left(q - 2 k_F^-\right)\Biggr).
\end{aligned}
\end{equation}

The scattering rates $1/\tau_{\pm}$ at the Fermi energies are given by, for $\alpha = \pm$,

\begin{equation}
\begin{aligned}
        \frac{1}{\tau_{\alpha}} = \frac{4 m n_{LR}}{\pi \hbar^3} \int_0^{\pi/2} & \text{d} \theta \, \sin^2{\left(\theta\right)} \\
        &\left(\frac{\frac{2 \pi e^2}{2 k_{F}^\alpha \kappa \sin{\left(\theta\right)}}}{1 + \Pi_0 \left(2 k_F^\alpha \sin{\left(\theta\right)}\right)\frac{2 \pi e^2}{2 k_{F}^\alpha \kappa \sin{\left(\theta\right)}}}\right)^2.
\end{aligned}
\end{equation}

Therefore, the spin averaged scattering time is the following.

\begin{equation}
\begin{aligned}
\tau
={}&
\frac{\pi\hbar^3}{8mn_{LR}}
\sum_{\alpha=\pm}
\left(1-\alpha\frac{B}{B_c}\right)
\\
&\times
\left[
\int_0^{\pi/2}\mathrm{d}\theta\,
\sin^2\theta
\left(
\frac{
\dfrac{2\pi e^2}
{2k_F^\alpha\kappa\sin\theta}
}{
1+
\Pi_0\left(2k_F^\alpha\sin\theta\right)
\dfrac{2\pi e^2}
{2k_F^\alpha\kappa\sin\theta}
}
\right)^2
\right]^{-1}
\end{aligned}
\end{equation}

Therefore, the conductivity is

\begin{equation} \label{eq:sigma_B}
\begin{aligned}
\sigma = {}&
\frac{\pi \hbar^3n e^2}{8 m^2 n_{LR}}
\sum_{\alpha=\pm}
\left(1-\alpha\frac{B}{B_c}\right)
\\
&\times
\left[
\int_0^{\pi/2}\mathrm{d}\theta\,
\sin^2\theta
\left(
\frac{
\dfrac{2\pi e^2}
{2k_F^\alpha\kappa\sin\theta}
}{
1+
\Pi_0\left(2k_F^\alpha\sin\theta\right)
\dfrac{2\pi e^2}
{2k_F^\alpha\kappa\sin\theta}
}
\right)^2
\right]^{-1}
\end{aligned}
\end{equation}

We set this equal to the conductivity obtained by considering a system with the same disorder parameters and $n$ as before, but with $B = 0$ and total degeneracy $g_\text{eff}$. The conductivity in that case is given by

\begin{equation} \label{eq:sigma_eff}
    \sigma = \frac{\pi \hbar^3n e^2}{4 m^2 n_{LR}}  \left(\int_0^{\pi/2} \text{d} \theta \, \sin^2{\left(\theta\right)} \left(\frac{\frac{2 \pi e^2}{2 \sqrt{\frac{4 \pi n}{g_\text{eff}}} \kappa}}{\sin{\left(\theta\right)} + \frac{g_\text{eff} m e^2}{2 \sqrt{\frac{4 \pi n}{g_\text{eff}}} \kappa \hbar^2}}\right)^2\right)^{-1}
\end{equation}

Setting $\sigma$ equal from (\ref{eq:sigma_B}) and (\ref{eq:sigma_eff}), we obtain the following equation for $g_\text{eff}$, which can be used to give $g_\text{eff}$ numerically.

\begin{equation} \label{eq:geff}
\begin{aligned} 
    &\frac{1}{2}
    \sum_{\alpha=\pm}
    \left(1-\alpha\frac{B}{B_c}\right)
    \\
    &\times
    \left[
    \int_0^{\pi/2}\mathrm{d}\theta\,
    \sin^2\theta
    \left(
    \frac{
    \dfrac{2\pi e^2}
    {2k_F^\alpha\kappa\sin\theta}
    }{
    1+
    \Pi_0\left(2k_F^\alpha\sin\theta\right)
    \dfrac{2\pi e^2}
    {2k_F^\alpha\kappa\sin\theta}
    }
    \right)^2
    \right]^{-1}\\
    & = \left(\int_0^{\pi/2} \text{d} \theta \, \sin^2{\left(\theta\right)} \left(\frac{\frac{2 \pi e^2}{2 \sqrt{\frac{4 \pi n}{g_\text{eff}}} \kappa}}{\sin{\left(\theta\right)} + \frac{g_\text{eff} m e^2}{2 \sqrt{\frac{4 \pi n}{g_\text{eff}}} \kappa \hbar^2}}\right)^2\right)^{-1}.
\end{aligned}
\end{equation}

The RHS of (\ref{eq:geff}) can be simplified in the case of strong screening ($q_{TF}/\left(2 k_F\right) \gg 1$, with $q_{TF}$ being the 2D Thomas-Fermi screening constant) to yield a simpler expression for $g_\text{eff}$.

\begin{equation}
\begin{aligned}
&g_{\mathrm{eff}}
=
\left[
\frac{\pi^3\hbar^4}{2m^2}
\sum_{\alpha=\pm}
\left(1-\alpha\frac{B}{B_c}\right)
\right.
\\
&\left.\!
\left\{
\int_0^{\pi/2}\mathrm{d}\theta\,
\sin^2\theta
\left(
\frac{
\dfrac{2\pi e^2}
{2k_F^\alpha\kappa\sin\theta}
}{
1+
\Pi\left(2k_F^\alpha\sin\theta\right)
\dfrac{2\pi e^2}
{2k_F^\alpha\kappa\sin\theta}
}
\right)^2
\right\}^{-1}
\right]^{1/2}
\end{aligned}
\end{equation}

The case of strong screening is not entirely an unjustified assumption as most MITs happen at low charge carrier densities, and this condition holds for low enough charge carrier densities, where $q_{TF} \gg k_F$, by virtue of $k_F \sim n^{1/2}$ and $q_{TF}$ being density-independent in 2D. The equations for $g_\text{eff}$ so far only consider in-plane charged scattering, but we can easily include out-of-plane charged scatterers and short-range scatterers by simply equating the conductivities obtained by the two methods as in the above prescription.

\begin{figure*}[t]
    \centering
    \includegraphics[width=\textwidth]{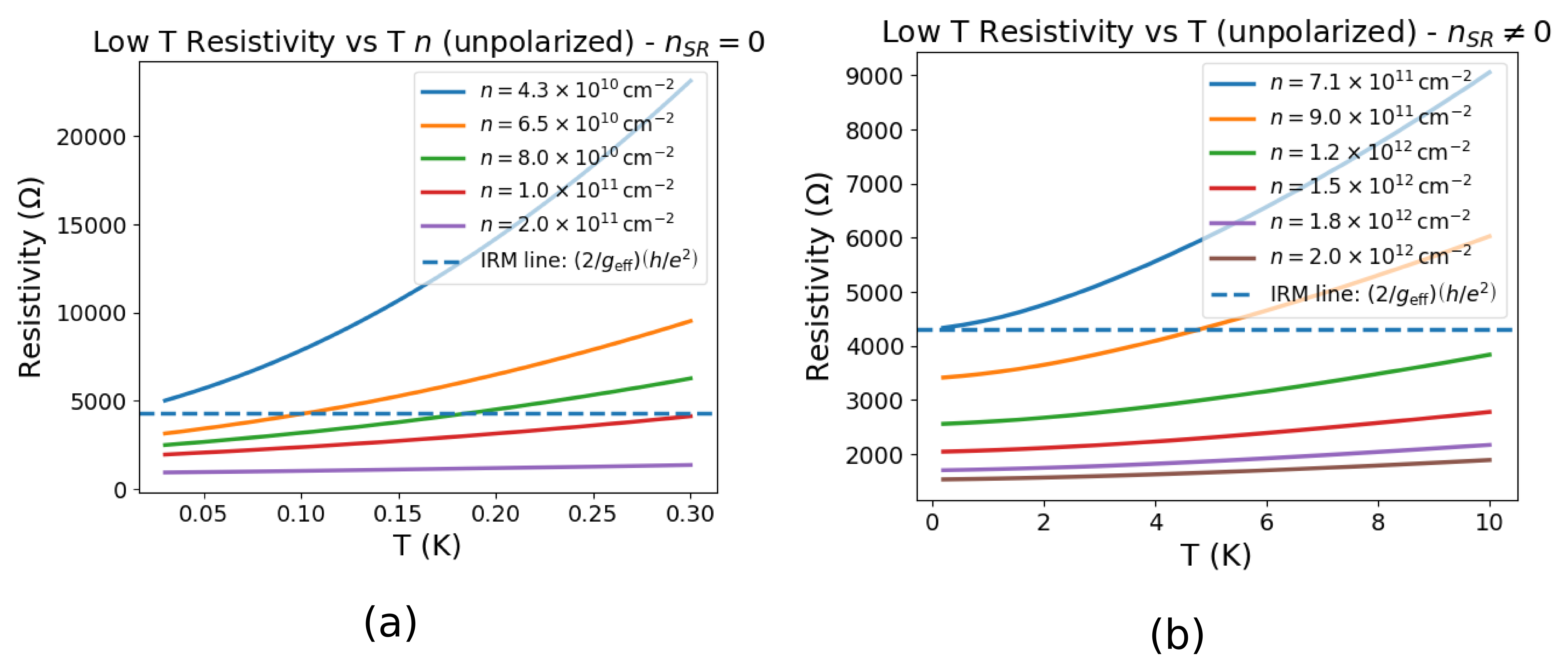}
    \caption{$B = 0$ resistivities for varying $n$ \textbf{(a)} Resistivity at $B = 0$ ($n_{SR} = 0$) and \textbf{(b)} Resistivity at $B = 0$ ($n_{SR} \neq 0$).}
    \label{fig:unpolarizedrhos}
\end{figure*}

\begin{figure*}[t]
    \centering
    \includegraphics[width=\textwidth]{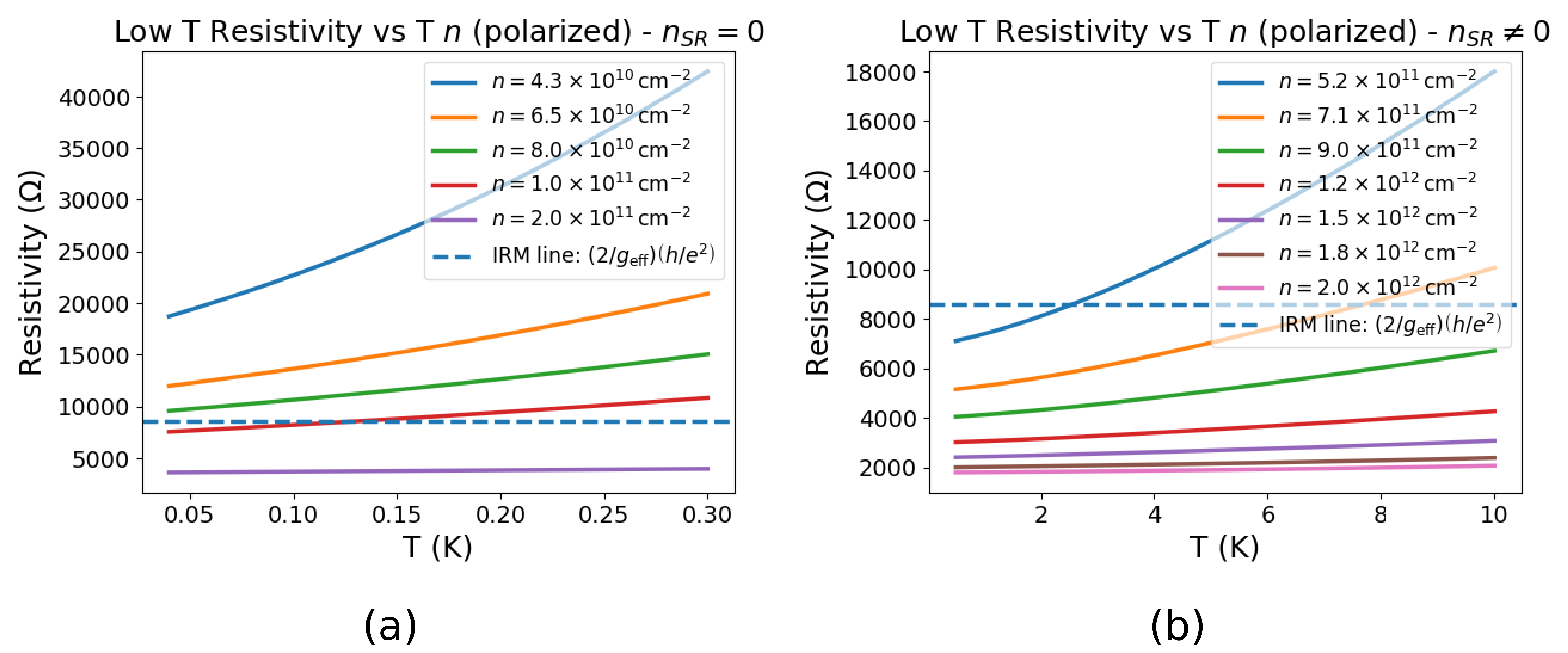}
    \caption{Resistivities for varying $n$  when $B$ is just enough to polarize all $n$ at $T = 0$ \textbf{(a)} Resistivity when $B$ is just enough to polarize all $n$ at $T = 0$ ($n_{SR} = 0$) and \textbf{(b)} Resistivity when $B$ is just enough to polarize all $n$ at $T = 0$ ($n_{SR} \neq 0$).}
    \label{fig:polarizedrhos}
\end{figure*}

For the parameters of \cite{ge2025visualizing, haleemprivatecomm}, we plot the resistivity at low temperatures for various values of $n$ at $B = 0$, where in Figure \ref{fig:unpolarizedrhos}a, we have set the short range disorder density $n_{SR} = 0$, while the long range disorder takes the value it did in the paper \cite{ge2025visualizing} $\left(n_{LR} = 1.6 \times 10^{11} \text{ cm}^{-2}\right)$. In Figure \ref{fig:unpolarizedrhos}b, we have also used the experimental value of $n_{SR}$ $\left(n_{SR} = 3.5 \times 10^{11} \text{ cm}^{-2}\right)$. For the highest $n$ in each figure, we use a large enough magnetic field $\sim 2.55 \, \text{T}$ (in Figure \ref{fig:polarizedrhos}a) and $\sim 25.5 \, \text{T}$ (in Figure \ref{fig:polarizedrhos}b) to spin polarize the system at $T = 0$, and for that value of $B$, we plot the corresponding Figures \ref{fig:polarizedrhos}a and \ref{fig:polarizedrhos}b.

\begin{figure*}[t]
    \centering
    \includegraphics[width=\textwidth]{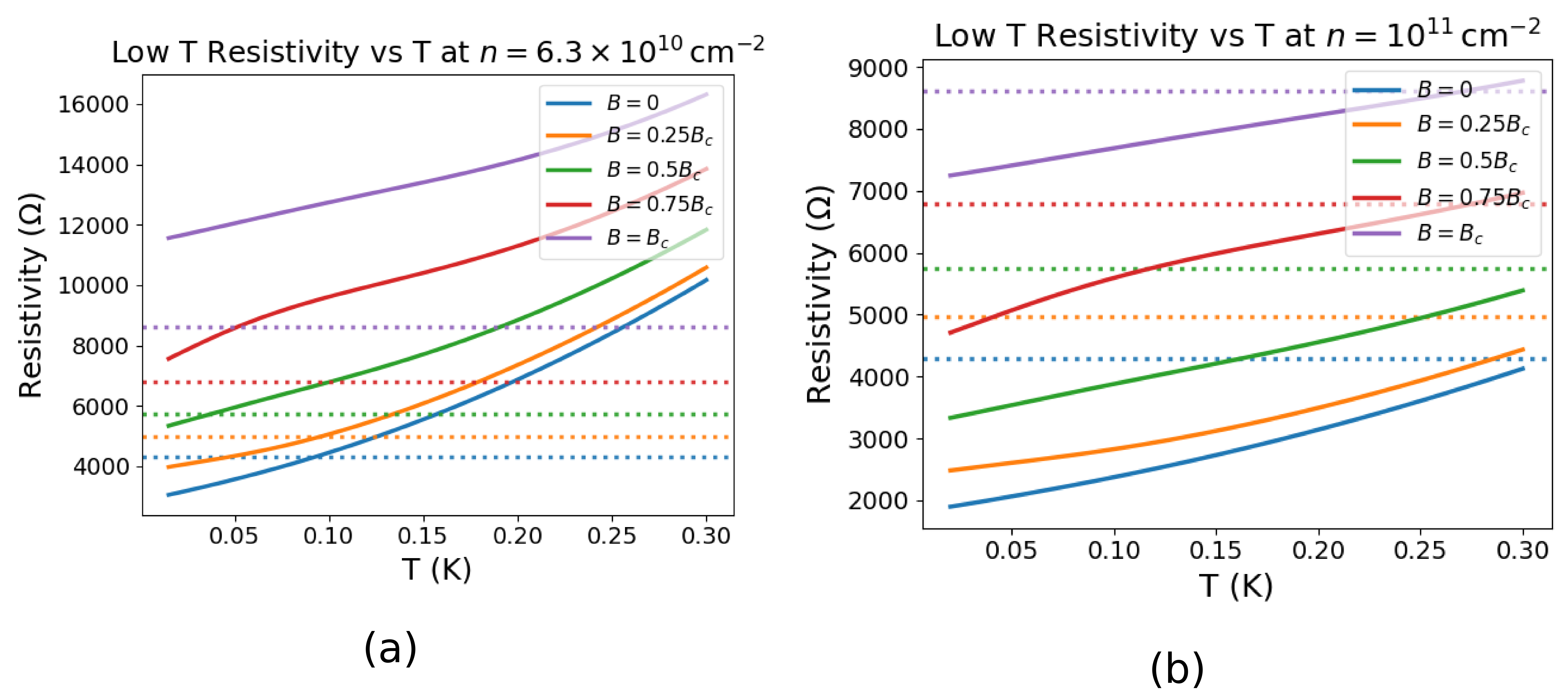}
    \caption{$\rho$ for $B = 0, B_c/4, B_c/2, 3B_c/4, B_c$ for \textbf{(a)} $n = 6.3 \times 10^{10} \text{ cm}^{-2}$ \textbf{(b)} $n = 10^{11} \text{ cm}^{-2}$, with the corresponding IRM criterion for each $B$. The solid curves represent $\rho (T)$ for each $B$ and the dotted lines of the same color give the corresponding IRM criterion $\left(\rho = \left(h/e^2\right)\left(2/g_{\text{eff}}\right)\right)$.}
    \label{fig:Bvaryingrhos}
\end{figure*}

For $n = 6.3 \times 10^{10} \text{ cm}^{-2}$ and $n = 10^{11} \text{ cm}^{-2}$, we call the minimum magnetic field that polarizes entirely at $T = 0$ $B_c$. We plot the resistivity at low $T$ for $B = 0, B_c/4 ,B_c/2, 3B_c/4, B_c$ for $n_{LR}$ the same as in the Berkeley experiment and $n_{SR} = 0$, along with the corresponding IRM lines for different $g_\text{eff}$ values on the same graph, since the system is not fully unpolarized or polarized for intermediate $B$ values (Figures \ref{fig:Bvaryingrhos}a and \ref{fig:Bvaryingrhos}b).

For $T = 0$, we also derive an asymptotic expression for $\rho (T = 0, B)$ in the limit $B \ll B_c$. We define $x=B/B_c$ and, for now, align $B$ along the spin direction that we label $+$, so that $x>0$. Therefore, at $T=0$ and for $x\ll 1$, we have $n_+=n(1-x)/2$ and $n_-=n(1+x)/2$. Note that the labels $+$ and $-$ are interchangeable, so we expect our results to remain unchanged under $x\leftrightarrow -x$. This means that $x$ can only enter our expression through $|x|$. The zero-field Fermi wave vector is $k_F=\sqrt{2\pi n/g_v}$, and we define $s=q_{TF}/\left(2 k_F\right) = (2g_v)me^2/\left(\kappa\hbar^2(2k_F)\right)$. Our expression applies only to in-plane charged disorder with density $n_i$, as is appropriate for clean semiconductor samples.

We use the following expressions for our $x \ll 1$ asymptotic.

\begin{align}
    J_{1, 0} \left(s\right) &= \int_0^{\pi/2} \text{d} \theta \, \frac{s^2 \sin^2 {\left(\theta\right)}}{\left(\sin{\left(\theta\right) + s}\right)^2} \nonumber \\
    J_{1, 1} \left(s\right) &= \int_0^{\pi/2} \text{d} \theta \, \frac{s^2 \sin^3 {\left(\theta\right)}}{\left(\sin{\left(\theta\right) + s}\right)^3} \nonumber \\
    J_{1, 2} \left(s\right) &= \int_0^{\pi/2} \text{d} \theta \, \frac{s^2 \sin^3 {\left(\theta\right)} \left(s + 4 \sin{\left(\theta\right)}\right)}{4\left(\sin{\left(\theta\right) + s}\right)^4} \nonumber \\
    J_{2, 1} \left(s\right) &= \frac{\pi}{2}\frac{s^3}{\left(1+s\right)^3} \nonumber \\
    J_{2, \frac{3}{2}} \left(s\right) &= \sqrt{2}\frac{s^4}{\left(1+s\right)^4} \nonumber \\
    J_{2, 2} \left(s\right) &= \frac{\pi}{8}\frac{s^3 \left(s^2 + 11s + 7\right)}{\left(1+s\right)^5} \nonumber
\end{align}

With these expressions, we get for $B/B_c = x \left(\ll 1\right)$,

\begin{align}
    &\rho \left(T = 0, x\right) = \frac{h}{e^2} \frac{n_i}{n} \frac{2}{g_v^2} 
    \Bigg(J_{1, 0} \left(s\right) \nonumber \\
    &+ \frac{1}{2} \left(J_{2, 1} \left(s\right) - 2 J_{1, 1} \left(s\right)\right)|x| + \frac{1}{2} J_{2, \frac{3}{2}} \left(s\right) |x|^{3/2} \nonumber \\
    &+ \left(J_{1, 2} \left(s\right) + \frac{1}{2} J_{2, 1} \left(s\right) - \frac{1}{2} J_{2, 2}  \left(s\right) - \frac{1}{4} \frac{\left(J_{2, 1} \left(s\right)\right)^2}{J_{1, 0} \left(s\right)}\right) x^2 \nonumber \\
    &+ \mathcal{O} \left(|x|^{5/2}\right)\Bigg).
\end{align}

\begin{figure}
    \centering
    \includegraphics[width=0.9\linewidth]{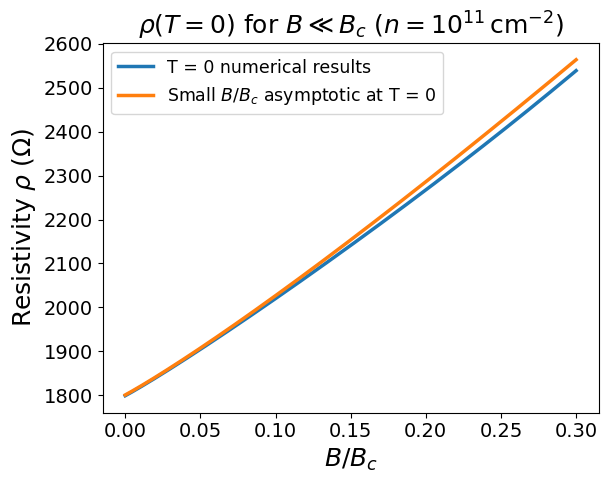}
    \caption{$T = 0$ numerical $\rho \left(B\right)$ for $B \ll B_c$ compared with the $B/B_c \ll 1$ asymptotic for $n = 10^{11} \text{ cm}^{-2}$ and $n_{LR} = 1.6 \times 10^{11} \text{ cm}^{-2}$.}
    \label{fig:lowBasymptotic}
\end{figure}

\begin{figure}
    \centering
    \includegraphics[width=0.9\linewidth]{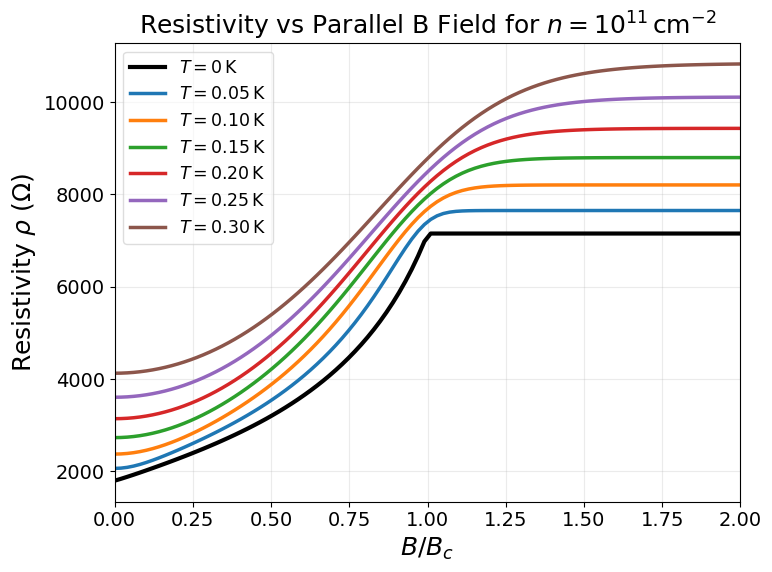}
    \caption{$\rho \left(B\right)$ for various fixed low $T$ for $n = 10^{11} \text{ cm}^{-2}$ and $n_{LR} = 1.6 \times 10^{11} \text{ cm}^{-2}$.}
    \label{fig:Bdependent}
\end{figure}

We plot the numerically obtained $\rho \left(T = 0\right)$ alongside this asymptotic result in Figure~\ref{fig:lowBasymptotic} for $n=10^{11} \text{ cm}^{-2}$. We also plot $\rho$ as a function of $B/B_c$ at various fixed low temperatures for the same density in Figure~\ref{fig:Bdependent}. In both figures, the short-range disorder density is zero, and the charged disorder density is $n_{\mathrm{LR}}=1.6\times10^{11}\text{ cm}^{-2}$.

\section{The strong screening limit for Coulomb disorder}

The strong-screening limit holds when $s=q_{TF}/(2k_F) \gg 1$. This condition is satisfied at low charge carrier density, since the Thomas-Fermi wave vector $q_{TF}$ is density-independent, whereas $k_F$ decreases as the charge carrier density is reduced. For a 2D system with charge carrier density $n$, and charged Coulomb in-plane impurities with density $n_{LR}$, the transport scattering rate $1/\tau_t$ is given by

\begin{equation}
    \frac{1}{\tau_t} = \frac{\hbar n_{LR}}{m} \left(\frac{2 \pi}{g} \right)^2 \, \frac{2}{\pi}\int_0^{\frac{\pi}{2}} \text{d} \theta \, \frac{2 s^2 \sin^2{\left(\theta\right)}}{\left(s + \sin{\left(\theta\right)}\right)^2}.
\end{equation}

The resistivity

\begin{equation}
    \rho = \frac{m}{n e^2 \tau_t} = \frac{h}{e^2}\frac{n_{LR}}{n} \frac{2 \pi}{g^2} \, \frac{2}{\pi}\int_0^{\frac{\pi}{2}} \text{d} \theta \, \frac{2 s^2 \sin^2{\left(\theta\right)}}{\left(s + \sin{\left(\theta\right)}\right)^2}.
\end{equation}

In order to calculate the density $n_\text{IRM}$ at which the IRM criterion is met, we set $\rho = h/e^2 \left(2/g\right)$, which gives

\begin{equation}
    \frac{\pi}{g} \, \frac{2}{\pi}\int_0^{\frac{\pi}{2}} \text{d} \theta \, \frac{2 s_\text{IRM}^2 \sin^2{\left(\theta\right)}}{\left(s_\text{IRM} + \sin{\left(\theta\right)}\right)^2} n_{LR} = n_\text{IRM},
\end{equation}

where $s_\text{IRM}$ is the value of $s$ at $n_\text{IRM}$. In the strong screening limit, we assume that the $s$ values accessed in the experiment, including $s_\text{IRM}$, are $\gg 1$, therefore in the strong screening limit,

\begin{equation}
    \frac{\pi}{g} \, n_{LR} = n_\text{IRM}.
\end{equation}

This implies that the critical carrier density at the MIT is parametrically given by the charged impurity density, but with an $\mathcal{O} \left(1\right)$ pre-factor of $\pi/g$, where $g$ is the total degeneracy including spin and valley. We now compute $n_{\mathrm{IRM}}$ in the strong-screening limit for the following systems, using typical experimental disorder densities and material parameters. We assume the charged impurity density $n_{LR} = 10^{10} \text{cm}^{-2}$ to be the same in all the cases below, except for the BL-MoSe$_2$ system, for which we use the value of $n_{LR}$ in the Berkeley experiment~\cite{ge2025visualizing}.

\begin{itemize}
    \item Graphene electrons with $m = 0.033 m_e$, $g_s g_v = 2 \times 2 = 4$, $n_{LR} = 10^{10} \text{cm}^{-2}$, $\kappa = 4$.
    \item Graphene holes with $m = 0.6 m_e$, $g_s g_v = 2 \times 2 = 4$, $n_{LR} = 10^{10} \text{cm}^{-2}$, $\kappa = 4$.
    \item GaAs electrons with $m = 0.07 m_e$, $g_s g_v = 2 \times 1 = 2$, $n_{LR} = 10^{10} \text{cm}^{-2}$, $\kappa = 10.5$.
    \item GaAs holes with $m = 0.5 m_e$, $g_s g_v = 2 \times 1 = 2$, $n_{LR} = 10^{10} \text{cm}^{-2}$, $\kappa = 10.5$.
    \item BL-MoSe$_2$ electrons with $m = 0.54 m_e$, $g_s g_v = 2 \times 6 = 12$, $n_{LR} = 1.6 \times 10^{11} \text{cm}^{-2}$, $\kappa = 3.73$.
\end{itemize}

We compute the $n_\text{IRM}$ for the above 5 systems assuming strong screening, and later use $n_\text{IRM}$ to find $s_\text{IRM}$ to verify if the strong screening approximation is valid. We then compute the ratio $n_\text{WC}/n_\text{IRM}$ to compare the densities at which the WC and IRM criteria are satisfied. These quantities are tabulated in Table \ref{tab:material_parameters}. As can be seen from the table, the above cases satisfy the strong screening condition, and for all these ultra clean systems, except for the graphene electrons and GaAs electrons, we expect the WC criterion to happen at higher density than the strong localization condition. Of course, if the impurity density is increased, then localization would happen at higher densities, possibly modifying this conclusion.

In the strong field limit, where all the carriers are spin polarized, the degeneracy decreases by a factor of 2, and consequently, for fixed $n$, $s$ decreases by a factor of $2^{3/2}$ and the IRM condition changes by a factor of 2, leading to an increase in $n_\text{IRM}$.

\begin{table*}[t]
\caption{
Material parameters, disorder densities, and corresponding IRM and WC related quantities.
}
\label{tab:material_parameters}
\begin{ruledtabular}
\begin{tabular}{lcccccccc}
Material and charge carriers
& $m/m_e$
& $\kappa$
& $g$
& $n_{\mathrm{LR}} \left(\text{cm}^{-2}\right)$
& $n_{\mathrm{IRM}} \left(\text{cm}^{-2}\right)$
& $s_{\mathrm{IRM}}$
& $n_\text{WC}/n_\text{IRM}$
\\
\hline
Graphene electrons & 0.033 & 4 & 4 & $10^{10}$ & $7.85 \times 10^9$ & 20 & $7.3 \times 10^{-2}$ \\
Graphene holes & 0.6 & 4 & 4 & $10^{10}$ & $7.85 \times 10^9$ & 363 & 24 \\
GaAs electrons & 0.07 & 10.5 & 2 & $10^{10}$ & $1.57 \times 10^{10}$ & 4.04 & $2.4 \times 10^{-2}$ \\
GaAs holes & 0.5 & 10.5 & 2 & $10^{10}$ & $1.57 \times 10^{10}$ & 29 & 1.2 \\
BL-MoSe$_2$ electrons & 0.54 & 3.73 & 12 & $1.6 \times 10^{11}$ & $4.19 \times 10^{10}$ & 789 & 4.2 \\
\end{tabular}
\end{ruledtabular}
\end{table*}

\section{Discussion in the context of the Berkeley experiment}

In~\cite{babbar2026wignersolidandersonsolid}, we argue that the MIT observed in the experiment~\cite{ge2025visualizing, haleemprivatecomm} is more consistent with the IRM criterion for strong localization than it is with a WC transition. As argued in the paper~\cite{babbar2026wignersolidandersonsolid}, for all values of $n$ studied in the experiment, $q_{TF}/\left(2 k_F\right) \gg 1$, for the unpolarized system $g_v = 6, g_s = 2$. This criterion will continue to hold when the system is fully polarized $g_v = 6, g_s = 1$. Therefore, the resistivity at $T = 0$ for the range of densities they probe, for short range disorder density $n_{SR}$, with the parameter $\left(m V_0/\hbar^2\right)^2 \simeq 2$ (as argued in~\cite{babbar2026wignersolidandersonsolid}) giving the strength of this disorder, and charged (long-range) disorder density $n_{LR}$ is given by

\begin{equation} \label{eq:rhoeq}
    \rho \simeq \frac{h}{e^2}\frac{n_{LR} \left(\frac{2 \pi}{g}\right)^2 + 2 n_{SR}}{2 \pi n}.
\end{equation}

For the paper's low disorder density regime (LDD), they only go up to $n = 2.55 \times 10^{12} \text{ cm}^{-2}$. If we consider $n < 2 \times 10^{12} \text{ cm}^{-2}$, then we need fields up to $25.5 \text{ T}$ to polarize all values of $n$ at $T = 0$, with equality for $n = 2 \times 10^{12} \text{ cm}^{-2}$. For a general $n$, the value needed to polarize is given by 

\begin{equation}
    B_c = \frac{2 \pi \hbar^2 n}{g_v m g_L \mu_B}.
\end{equation}

\begin{figure*}
    \centering
    \includegraphics[width=\textwidth]{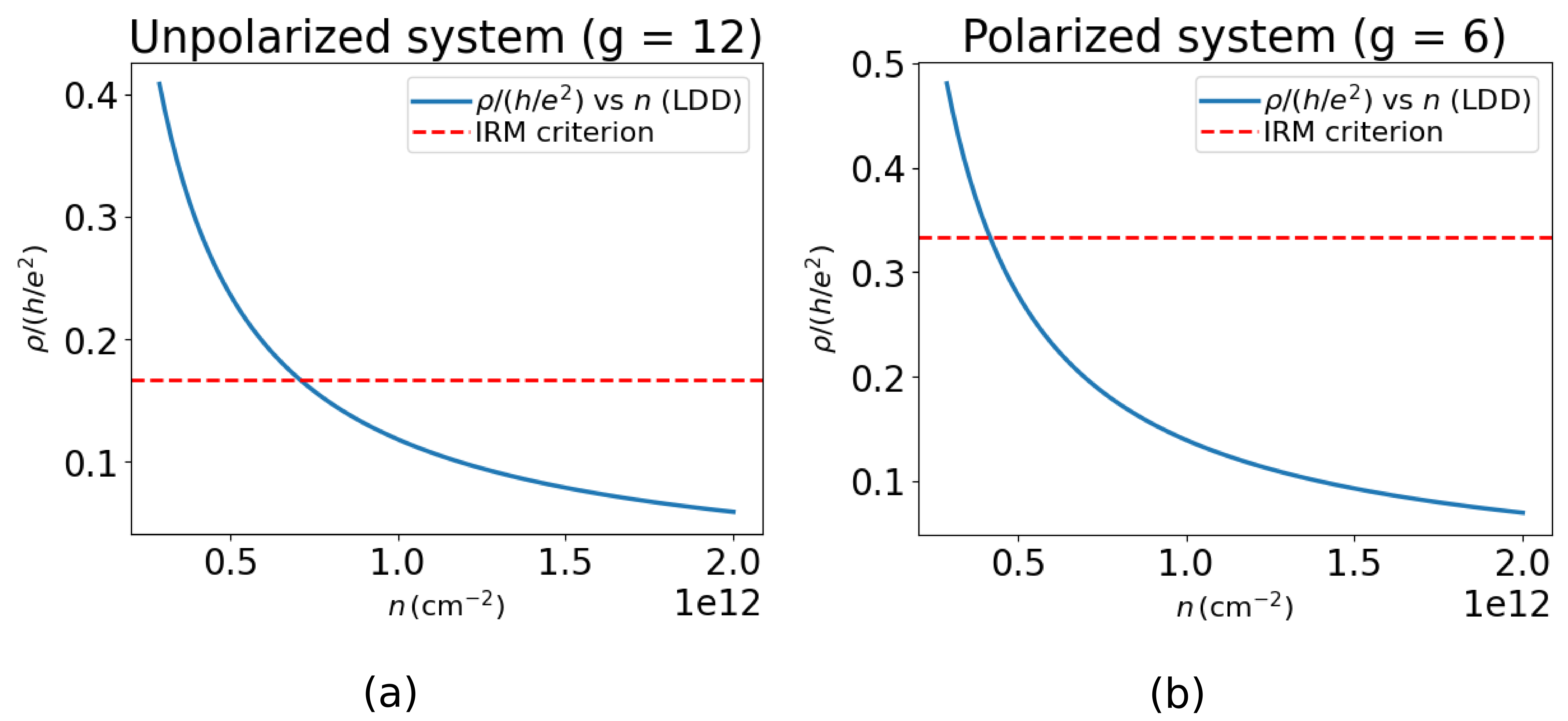}
    \caption{$\rho \left(T = 0\right)$ for long-ranged and short-ranged disorder as in the Berkeley experiment \textbf{(a)} unpolarized \textbf{(b)} polarized.}
    \label{fig:nsrnot0T0}
\end{figure*}

The resistivities at $T = 0$ for varying values of $n$ are given, under the RPA-Boltzmann approximation for the unpolarized case by Figure \ref{fig:nsrnot0T0}a. If we fully spin-polarize the system, then the resistivities are given by Figure \ref{fig:nsrnot0T0}b. In both graphs, the line for the IRM criterion is also drawn. As the dominant contributor is short-ranged disorder, the resistivity is roughly unchanged. The IRM criterion is $\rho_c = \left(2/g\right) \left(h/e^2\right)$, and $\rho_c$ increases by a factor of $2$. As $\rho \propto 1/n$, $n_c$ halves when polarized, and if the transition is due to strong localization, this should be observed. Note that this halving the critical density here is entirely a result of the disorder being short-ranged in the Berkeley experiment so that there is no spin polarization induced screening effect operational in transport. The effect here arises entirely from the increase in $k_F$ induced by spin polarization, which increases the IRM resistivity $\left(\rho_c\right)$. The actual resistivity of the system remains unchanged.

If we were to repeat this calculation by setting again at $T = 0$, $n_{SR} = 0$, i.e. now all scattering is by screened long-range Coulomb scatterers, then for the unpolarized case, we get the following Figure \ref{fig:nsr0T0}a and for the polarized case, we get the following Figure \ref{fig:nsr0T0}b ($n$ values are adjusted to be close to $n_c$). As can be seen from the equation (\ref{eq:rhoeq}), the resistivity in the metallic range (where our equation applies) increases by a factor of $4$ when polarized for the same $n$. The IRM criterion is $\rho_c = \left(2/g\right) \left(h/e^2\right)$, and $\rho_c$ also increases by a factor of $2$. As $\rho \propto 1/n$, $n_c$ increases by a factor of $2$ when polarized, and if the transition is due to strong localization, this should be observed.

Thus, the physics is very different depending on whether the disorder scattering is intrinsically short-ranged defect scattering or long-ranged Coulomb scattering: in the first case, $n_c$ decreases and in the second case $n_c$ increases!

We note that, for particular relative strengths of short- and long-range disorder, their opposing effects on the critical density $n_c$ may cancel. We consider a regime similar to that of the Berkeley experiment, assuming strong screening and denoting the short-range disorder strength $\left(mV_0/\hbar^2\right)^2$ by $S$. In the unpolarized case,
\[
n_c=\frac{g_v}{2\pi}\left(n_{\rm LR}\left(\frac{\pi}{g_v}\right)^2+S n_{\rm SR}\right),
\]
whereas in the fully polarized case,
\[
n_c=\frac{g_v}{4\pi}\left(n_{\rm LR}\left(\frac{2\pi}{g_v}\right)^2+S n_{\rm SR}\right).
\]
Equating these two expressions gives
\[
\frac{Sg_v^2}{2\pi^2}n_{\rm SR}
= n_{\rm LR}.
\]
Thus, when the short- and long-range disorder components satisfy this relation, their respective effects on $n_c$ exactly compensate. Consequently, within an IRM description of a disorder-driven MIT, the crossover density remains unchanged upon full spin polarization.

\begin{figure*}[t]
    \centering
    \includegraphics[width=\textwidth]{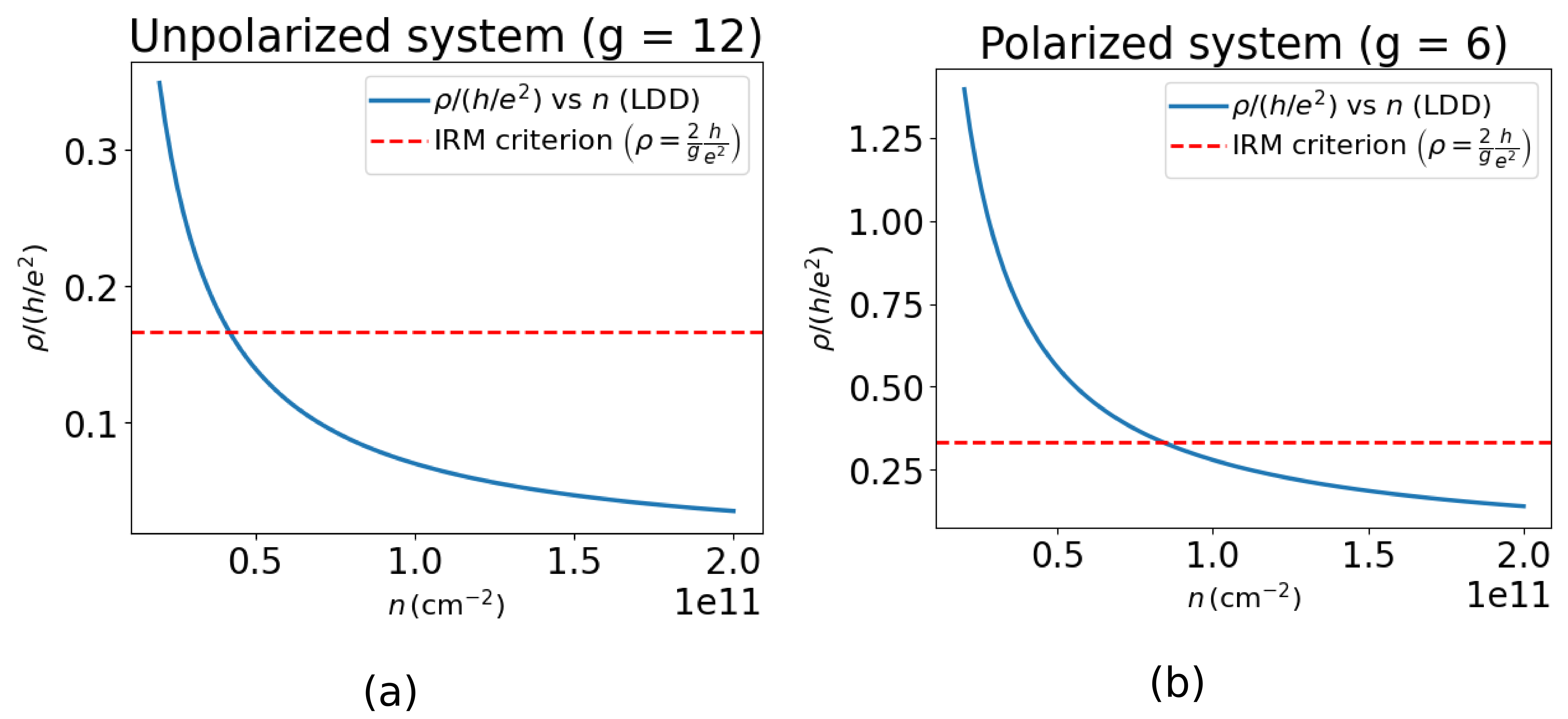}
    \caption{$\rho \left(T = 0\right)$ for only the long-ranged disorder as in the Berkeley experiment \textbf{(a)} unpolarized \textbf{(b)} polarized.}
    \label{fig:nsr0T0}
\end{figure*}

\section{Discussion in the context of the MIT experiment}

There was a recent transport experiment at MIT in which a metal-insulator transition was reported in rhombohedral graphene, and different regions in the phase diagram were interpreted as evidence for a Wigner crystal and a metallic Wigner crystal phase. In this section, we argue that the metal-insulator transition in the hexalayer rhombohedral graphene sample, assuming valley degeneracy $g_v=2$, can also be explained by strong localization: charge carriers move in an effective disorder potential generated by screened charged impurities.

\begin{figure*}[t]
    \centering
    \includegraphics[width=\textwidth]{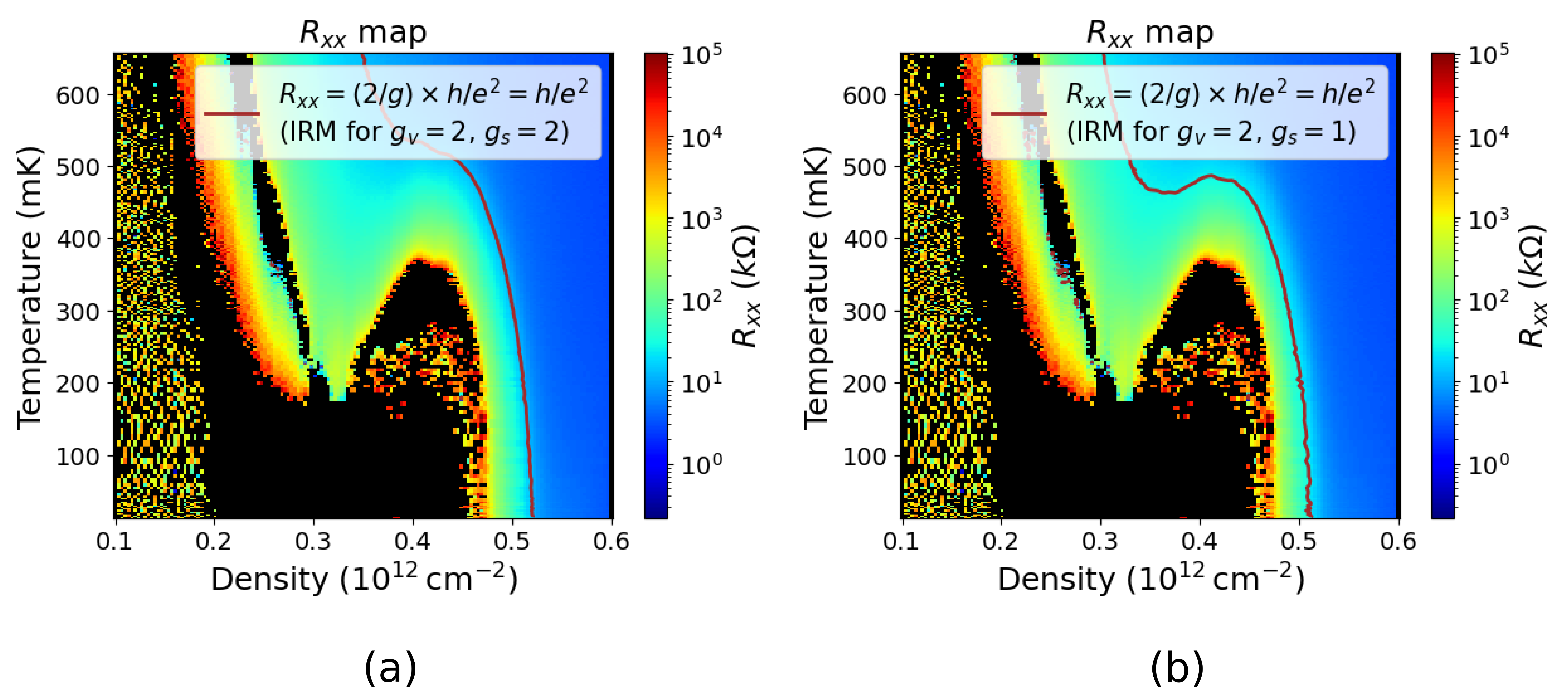}
    \caption{$R_{xx}$ against $T$ and $n$ (experimental data from the MIT experiment) with the IRM curve for the \textbf{(a)} unpolarized system \textbf{(b)} spin-polarized system drawn. Including the aspect ratio of the material may change the location of the IRM curve. See text for details.}
    \label{fig:longju}
\end{figure*}

In Figure \ref{fig:longju}, we show the experimental $R_{xx}$ data as a function of temperature and charge-carrier density. We overlay the resistivity corresponding to the IRM criterion,
$R_{xx} = \frac{2}{g}\frac{h}{e^2}$, in Figure \ref{fig:longju}a. We note that the IRM line is pretty close to the experimental phase boundary separating the metal from the insulator, thus making a suggestive case for the insulator being a strongly-localized system rather than a WC. We note, however, that the plots are for $R_{xx}$, not $\rho_{xx}$, and considering the aspect ratio of the sample, which we have assumed to be unity, to compute $\rho_{xx}$ may change the location of the IRM curve on the plot. Since the sample aspect ratio is not known, the IRM line's precise position is therefore subject to some uncertainty; however, this uncertainty does not affect the physical conclusions drawn from the comparison.

If the transition is really driven by strong localization in the screened impurity potential, then polarizing the system should change the total degeneracy $g$, thereby shifting the IRM criterion. We therefore redraw the corresponding IRM curve for the polarized case in Figure \ref{fig:longju}b. We skip the details here since they are already given above for the Berkeley experiment -- all that change are the relevant system parameters in considering graphene in the MIT experiment.

We do mention, however, that graphene being typically an exceptionally clean material, it is conceivable that the high-density part of the insulating phase is mostly WC physics, whereas the low-density part is certainly a disorder-driven localized phase. There is no phase transition between these two regimes -- it is a smooth crossover where the effective short range order (or equivalently, the Larkin length) increases from being less than a WC lattice constant at low density to being larger than the effective WC lattice constant at high density.

\section{Conclusion}

In this work, we have derived the resistivity of a sample with 2D charge carriers with a parabolic dispersion relation, in the presence of a parallel magnetic field, assuming the only effect of the field is to modify the energy of the two spins of the carriers. We argue that in transport experiments observing an MIT where the formation of a pinned WC is often claimed as the reason of the MIT, disorder plays an important role. More generally, the observation of a density-driven MIT, or of strongly insulating transport, does not by itself uniquely identify the microscopic nature of the insulating state. The crossover density at which a disorder-driven MIT is observed is a function of the degeneracy of the system, and if we apply a parallel field, this degeneracy can be changed, therefore changing the crossover density. On the contrary, for a pinned WC MIT, the crossover density is independent of the degeneracy, since the ferromagnetic and the antiferromagnetic WC have essentially the same energy in the WC phase by virtue of the exponentially suppressed exchange effect in a solid. Therefore, a parallel field can be used to distinguish between a disorder-driven and an interaction driven MIT. 

We also derive an equation for the effective spin and valley degeneracy $g_\text{eff}$, which inevitably enters the IRM criterion for strong localization. In addition, we obtain the low-field asymptotic form of the zero-temperature resistivity, $\rho(T=0,B\ll B_c)$. In the context of the experiments at Berkeley~\cite{ge2025visualizing, haleemprivatecomm} and MIT~\cite{longjumetallicwignercrystal}, we give quantitative predictions about the system when a parallel field is applied, allowing for a direct verification of our theory. In particular, we find that for a strong localization driven transition, for the disorder parameters the same as that of the experiment at Berkeley, when the in-plane long-range disorder is dominant, the crossover density doubles when the system is polarized. On the other hand, if the dominant contributor is short-ranged disorder, the crossover density halves when the system is polarized. This arises from the two competing effects of spin polarization: the effective Fermi momentum $k_F$ increases with spin polarization and the effective screening decreases because $q_{TF}/k_F$ decreases. Whether the localization transition density increases or decreases depends on whether the bare disorder in the system is long ranged (where screening is effective) or short ranged (where screening is ineffective).

However, it may also be the case that a WC is stabilized in the presence of a parallel field due to ferromagnetic coupling between neighboring charge carriers. This would also lead to the insulating phase persisting to higher densities, the same effect as if the dominant disorder contributor was charged disorder, but this effect should be very small since QMC calculations show the ferromagnetic WC to be the stable ground state, so a parallel field should have no observable effect~\cite{drummond2009phase}.

As an aside, we point out that existing density-tuned MIT experiments in the highest quality 2D samples support the experimental MIT as arising from electron localization rather than electron crystallization. As already mentioned above, the highest quality 2D GaAs hole samples of~\cite{manfraetal} show an MIT at a carrier density low enough so that the effective transition happens at $r_S \sim 50 \, (\gg r_S = r_\text{WC} \sim 37)$. This can be explained as a localization transition because in such a pure sample $n_\text{IRM}$ would be very low. There is also an experiment for 2D electrons in extreme high quality GaAs heterostructures~\cite{GaAs_heterostructure_1, GaAs_heterostructure_2}, where in spite of the extreme high purity, the observed MIT happens at an effective $r_S \sim 8$, which is much below $r_\text{WC} \sim 37$. This is because the 2D GaAs electrons have very small mass, making screening fairly ineffective (in contrast to 2D GaAs holes with large mass, where screening is very effective), thus making the effective disorder very strong even if the actual impurity density is low. Thus, both the unusually low $n_\text{MIT}$ for 2D GaAs holes and the unusually high $n_\text{MIT}$ for 2D GaAs electrons, both compared with the expected $n_\text{WC}$ for the two cases, are explicable using the transport theory based on screened Coulomb disorder and the IRM localization criterion. Neither experiment is compatible with a Wigner crystallization transition occurring at $r_S \sim 37$ since in one case (2D GaAs holes) the transition happens for $r_S \sim 50$ and in the other case (2D GaAs electrons) the MIT happens at $r_S \sim 8$. In this context, we also mention two recent Princeton experiments,  where direct spectroscopy and real space imaging by STM measurements show that the low-density 2D system transitions into a highly amorphous disordered localized phase characterized by the IRM criterion, and is not an electron crystal~\cite{Tsui2024WignerCrystal, aliyazdaniprivatecomm}.

In the context of screened charged disorder in the system, one immediate experimental consequence of the parallel field induced increase in $n_\text{IRM}$ as discussed in our work is that the measured parallel field induced magneto-resistance will be strongly enhanced already for $n > n_\text{MIT}$ at $B=0$. This is because, in a parallel field, $n_\text{MIT}$ will increase because of the increase in $n_\text{IRM}$ ($\sim n_\text{MIT}$ if the transition is disorder-induced), consequently driving the system into the localized insulating phase in the presence of a parallel field already at densities above the corresponding zero field value of $n_\text{MIT}$. Our theory cannot calculate this strongly enhanced magneto-resistance since the theory is explicitly for the metallic phase, but we can predict a giant magneto-resistance already at densities above (but close to) the zero field metal-to-insulator critical density based on our finding of the increase in $n_\text{IRM}$ associated with enhanced Coulomb scattering because of suppressed screening in the presence of the applied field.

Before concluding, we emphasize that the theory only includes the effects of the applied parallel field induced spin polarization and no orbital effects. So, it is important that the applied field is entirely in-plane with no out-of-plane component. Additionally, if the 2D layer has any finite thickness, then even a strong parallel field could affect orbital motions when the magnetic length is smaller than the thickness, which we do not consider in our work~\cite{parallel_field_earlier}. These orbital effects are expected to be small when the transverse thickness of the 2D layer is much smaller than the magnetic length, $l_B=\sqrt{\hbar/(eB)}$. In the context of the Berkeley experiment, the field required to fully polarize carrier densities $n<2\times10^{12}\,\mathrm{cm}^{-2}$ at $T=0$ is approximately $25.5$ T, which is already near the upper end of magnetic fields routinely accessible in 2D condensed-matter experiments. At this field, $l_B\simeq 50\,\text{\AA}$. Since the thickness of an individual MoSe$_2$ layer is of order $7\,\text{\AA}$~\cite{Park2018MoSe2Dielectric}, the bilayer thickness remains substantially smaller than the magnetic length even at these large fields. We therefore expect finite-thickness magneto-orbital effects to be relatively small in the regime considered here. 

Spin polarization affects the physics in two competing ways: (1) it modifies the Fermi surface by effectively increasing $k_F$ due to polarization; (2) it suppresses screening by reducing the density of states by suppressing the spin degeneracy. These two effects oppose each other in transport with increasing $k_F$ increasing the crossover resistivity set by the IRM criterion and decreasing screening enhancing the resistivity. Which of these two effects dominates depends on whether scattering from the short-ranged defect or long-ranged Coulomb impurities dominates transport. In the former case, resistivity does not change, so the effective MIT critical density is suppressed whereas in the latter case, the resistivity increases and the transition density increases with increasing magnetic field. In both cases, however, the applied parallel field should affect the 2D MIT transition if it is an Anderson localization transition, but the critical density should be unchanged if it is a Wigner crystal transition.

\section{Acknowledgments}

The authors gratefully acknowledge helpful communications on the experimental data with Long Ju of MIT and Haleem Kim of UC Berkeley. This work is supported by the Laboratory for Physical Sciences (LPS) through the Condensed Matter Theory Center (CMTC) at Maryland. AB thanks the Joint Quantum Institute at the University of Maryland for support through a JQI graduate fellowship.

\bibliography{references}

\end{document}